\documentclass[fleqn,10pt]{wlscirep}
\usepackage[utf8]{inputenc}
\usepackage[T1]{fontenc}
\title{Structural Evolution and Onset of the Density Wave Transition in the CDW Superconductor LaPt$_2$Si$_2$ Clarified with Synchrotron XRD}

\author[1,*]{E.~Nocerino}
\author[2]{I.~Sanlorenzo}
\author[3]{K.~Papadopulos}
\author[4]{M.~Medarde}
\author[4]{J.~Lyu}
\author[4]{Y.~M.~Klein}
\author[5]{A.~Minelli}
\author[6]{Z.~Hossain}
\author[7]{A.~Thamizhavel}
\author[2]{K.~Lefmann}
\author[8]{O.~Ivashko}
\author[8]{M.~v.~Zimmermann}
\author[3]{Y.~Sassa}
\author[1,$\dagger$]{M.~M{\aa}nsson}

\affil[1]{KTH Royal Institute of Technology, Department of Applied Physics, Alba Nova University Center, Stockholm, SE-114 21, Sweden}
\affil[2]{Nanoscience Center, Niels Bohr Institute, University of Copenhagen, Nørre Allé 59, 2100 Copenhagen, Denmark}
\affil[3]{Department of Physics, Chalmers University of Technology, SE-412 96 Göteborg, Sweden}
\affil[4]{Laboratory for Multiscale Materials Experiments, Paul Scherrer Institute, CH-5232 Villigen PSI, Switzerland}
\affil[5]{Inorganic Chemistry Laboratory, University of Oxford, Oxford OX1 3QR, United Kingdom}
\affil[6]{Department of Physics, Indian Institute of Technology, Kanpur 208016, India}
\affil[7]{DCMPMS, Tata Institute of Fundamental Research, Mumbai 400005, India}
\affil[8]{Deutsches Elektronen-Synchrotron DESY, Notkestr. 85, 22607 Hamburg, Germany}

\affil[*]{nocerino@kth.se}

\affil[+]{condmat@kth.se}

\begin{abstract}
The quasi-2D Pt-based rare earth intermetallic material LaPt$_2$Si$_2$ has attracted attention as it exhibits strong interplay between charge density wave (CDW) and and superconductivity (SC). However, the most of the results reported on this material
come from theoretical calculations, preliminary bulk investigations and powder samples, which makes it difficult to uniquely determine the temperature evolution of its crystal structure and, consequently, of its CDW transition. Therefore, the published literature around LaPt$_2$Si$_2$ is often controversial. In this paper, we clarify the complex evolution of the crystal structure, and the temperature dependence of the development of density wave transitions, in good quality LaPt$_2$Si$_2$ single crystals, with high resolution synchrotron X-ray diffraction data. According to our findings, on cooling from room temperature LaPt$_2$Si$_2$ undergoes a series of subtle structural transitions which can be summarised as follows: second order commensurate tetragonal ($P4/nmm$)-to-incommensurate structure followed by a first order incommensurate-to-commensurate orthorhombic ($Pmmn$) transition and then a first order commensurate orthorhombic ($Pmmn$)-to-commensurate tetragonal ($P4/nmm$). The structural transitions are accompanied by both incommensurate and commensurate superstructural distortions of the lattice. The observed behavior is compatible with discommensuration of the CDW in this material.
\end{abstract}
\begin{document}

\flushbottom
\maketitle
%
%
\thispagestyle{empty}


\section*{Introduction}

\begin{figure}[ht]
  \begin{center}
    \includegraphics[scale=0.5]{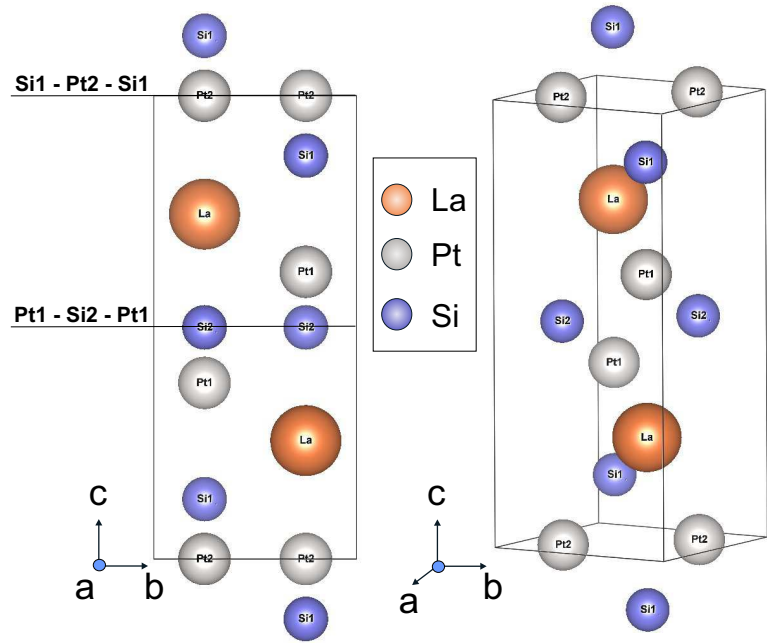}
  \end{center}
  \caption{Room temperature crystal structure of LaPt$_2$Si$_2$. The alternating Pt-Si planes are explicitly labelled.}
  \label{cell}
\end{figure}

A grand challenge in condensed matter physics is understanding the mechanisms underlying high temperature superconductivity (SC). Materials with competing electron spectrum instabilities, such as Cooper pairing and charge/spin-density waves (CDW/SDW), represent the ideal playground for this kind of investigations since the electron-phonon coupling established in such systems is believed to be a key factor in inducing SC. The quasi-2D Pt-based rare earth intermetallic material LaPt$_2$Si$_2$ belongs to this family of compounds as it exhibits strong interplay between CDW and SC. LaPt$_2$Si$_2$ crystallizes in a CaBe$_2$Ge$_2$-type tetragonal structure (space group $P4/nmm$), where two nonequivalent layers (Si1–Pt2–Si1) and (Pt1–Si2–Pt1) are arranged in alternating stacking separated by lanthanum atoms (Fig. \ref{cell}). In single-phased powder samples, indications of a first-order transition were observed from high temperature tetragonal to low temperature orthorhombic phase, accompanied by a CDW transition around T$_{CDW}$ = 112 K, where superlattice reflections corresponding to (n/3, 0, 0), with n = 1 and 2 were observed, followed by a SC transition at T$_c$ = 1.22 K \cite{gupta2016coexistence, nagano2013charge}. It was suggested that the CDW modulation of the crystal lattice would induce a tripling of the initial unit cell and that it would propagate in the P2 layer while superconductivity would be established in the P1 layer \cite{nagano2013charge}. The Fermi surface of LaPt$_2$Si$_2$ was found to have a two-dimensional nature \cite{hase2013electronic} and theoretical calculations of the phonon dispersion predict phonon-softening instabilities, leading to structural instabilities that are compatible with the (1/3, 0, 0) $q$-vector of the CDW. However, it was also shown that CDW and SC should coexist in the Pt1 layer \cite{hase2013electronic} and the softened phonon modes would mainly arise from Pt1. This finding suggested that the CDW transition occurs in the Pt1 layers, with large electron-phonon interaction \cite{kim2015mechanism}. The latter suggestion was later confirmed by Pt-NMR measurements \cite{aoyama2018195pt}, contrary to the aforementioned conjecture of Nagano \cite{nagano2013charge}. 

Beyond the results on polycrystalline samples \cite{kubo2014structural}, single crystal diffraction studies with in-house characterization methods were also reported. According to these studies the CDW transition occurs at 85 K (slightly lower than the 112 K reported from studies on polycrystalline samples) in correspondence to the maximum intensity of the superlattice satellites, which were found to have propagation vector (0.36 0 0), slightly different from the reported (1/3 0 0), indicating that the CDW modulation is incommensurate. Also, there was no clear indication of a structural transition towards orthorhombic crystal symmetry in single crystalline LaPt$_2$Si$_2$, showing that the properties of this material as a single crystal are slightly different from the ones as a polycrystal \cite{falkowski2019structural}.  Additionally, the CDW-induced modulation of the lattice has recently been found to develop a periodicity also along the $c$-axis with a different propagation vector and a different temperature dependence with respect to the previously identified one, propagating in the $ab$-plane \cite{falkowski2020multiple} which seems to indicate that multiple CDWs develop in this material. Finally, the superconductivity in this material, putatively non-conventional, (the Fermi surface exhibits 2 gaps of different magnitude according to $\mu^+$SR measurements \cite{das2018multigap}) was recently found to have SC gap and London penetration depth well described by standard BCS models \cite{nie2021nodeless}.

The published literature about the properties of this sample is slightly inconsistent and sometimes contradictory. At present, no experiments have measured the phonon dispersion and confirmed the theoretical predictions, nor provided direct evidence of the CDW transition temperature dependence, while the low temperature crystal structure in the CDW phase is not solved for powder nor for single crystalline LaPt$_2$Si$_2$. In this paper we present the results of a synchrotron XRD investigation on LaPt$_2$Si$_2$ single crystals, and clarify the complex evolution of the crystal structure in LaPt$_2$Si$_2$ as well as the temperature dependence and nature of its CDW state.

\section*{Results}

\subsection{\label{sec:xrd} Synchrotron XRD measurements}

\begin{figure*}[ht]
  \begin{center}
    \includegraphics[keepaspectratio=true,width=175 mm]{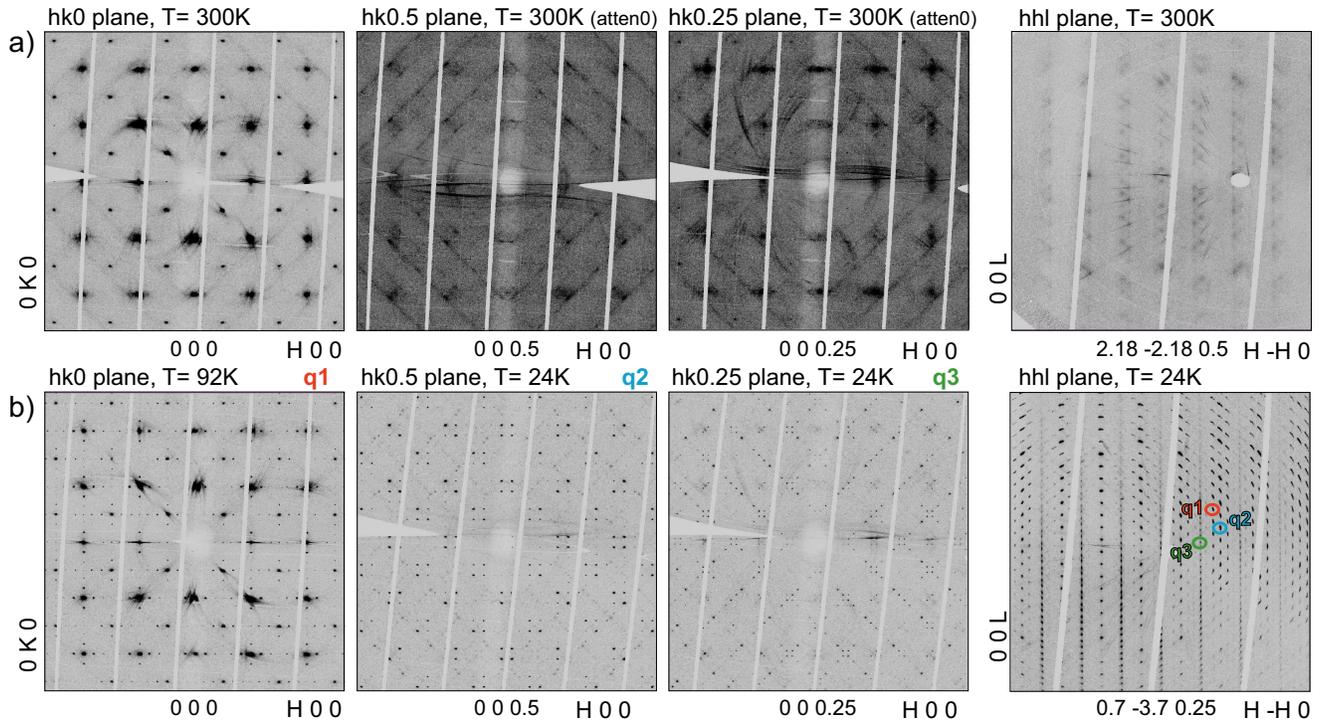}
  \end{center}
  \caption{Reciprocal space planes [$h$ $k$ $0$], [$h$ $k$ $0.5$], [$h$ $k$ $0.25$] and [$h$ $h$ $l$] at a) 300 K, showing the room temperature short-range diffuse scattering, and b) 92 K and 24 K, showing the 3 sets of satellites $q1$ (marked with the red color), $q2$ (marked in blue) and $q3$ (marked in green) as they appear. The first three cuts are therefore done by keeping the $hl$-plane constant, while shifting the $l$-axis in half-integer steps. The fourth cut is done along the $l$-axis in the $hk$ diagonal direction, to allow the simultaneous visualization of the three orders of satellites at 24 K. The origin of the different plots is marked at the bottom for each of them.}
  \label{planes}
\end{figure*}

Due to over-saturation of the structural Bragg peaks in the atten0 setting, the atten2 data were used for structural analysis and determination of the temperature dependence of the satellites. The atten0 data are presented in these paper for clarity of display of weak satellites and diffuse scattering. Figure \ref{planes} a) shows the $h$ $k$ $0$ (atten2), $h$ $k$ $0.5$ (atten0) and $h$ $k$ $0.5$ (atten0) planes in the reciprocal space at 300 K. Well structured diffuse scattering, indicating the presence of interactions with short-range correlation lengths induced by charge ordering, is already present at T = 300 K. Since the diffuse scattering does not converge on Bragg peaks at integer $h$ $k$ $l$, but it rather lies around them at semi-integer positions, the possibility of thermal diffuse scattering is excluded.

This implies that the charge ordering that results in the multi-$q$ modulation of the lattice is already in place in this temperature range. Therefore, the onset of the charge ordering in LaPt$_2$Si$_2$ is well above room temperature. The atten0 setting was chosen to display the $h$ $k$ $0.5$ and $h$ $k$ $0.25$ planes to highlight the presence of the diffuse scattering, since no structural Bragg peaks are allowed in this region of the reciprocal space and their over-saturation would not cause too much disturbance.
As the temperature decreases, we observe the appearance of satellites with 6 different propagation vectors: between 235 K and 35 K the satellites have average values of the propagation vectors $q\prime1$ = [0.35736 0 0] $\approx$ [0.36 0 0] and $q\prime \prime1$ = [0 0.35905 0] $\approx$ [0 0.36 0] in the full temperature range. Below 35 K the $q1-q’1$ satellites become weaker (but not enough to completely disappear) and the formation of the satellites $q\prime2$ = [0.18 0.18 0.5] and $q\prime \prime2$ = [0.18 -0.18 0.5] occurs. Looking at the reciprocal unit cell of the atten0 data at base temperature a third set of peaks could be identified with propagation vectors $q\prime3$ = [0.3 0.3 0.25] and $q\prime \prime3$ = [-0.3 -0.3 0.25]. This new set of satellites, whose intensity might possibly become stronger at lower temperature, could be noticed only by exploring the reciprocal cell in the zero attenuation data. Once they have been identified, appropriate cuts in the atten2 data allowed the observation of the $q3$ satellites in this dataset as well, despite the stronger attenuation conditions (Fig. \ref{planes} b)).

\begin{figure}[ht]
  \begin{center}
    \includegraphics[scale=0.45]{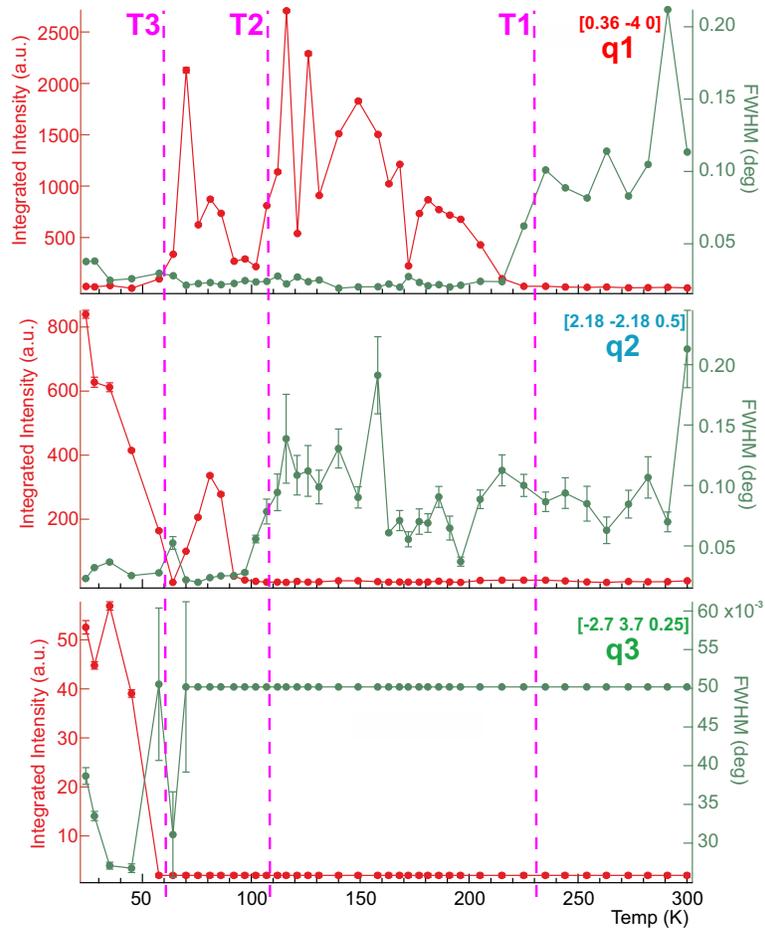}
  \end{center}
  \caption{Temperature dependence of the integrated intensities (left y axis) and FWHM (right y axis) for 3 satellites belonging to the 3 modulations of the lattice $q1$, $q2$ and $q3$, as for the labels on each panel. The transition temperatures T1, T2 and T3 are explicitly marked in magenta.}
  \label{temp_satellites}
\end{figure}

The reciprocal space planes $h$ $k$ $l$ are obtained by reconstructing the diffraction data in a single layer defined by the L1 = (1 0 0) and L2 = (0 1 0) vectors with the origin in (0 0 0), (0 0 0.5) and (0 0 0.25) respectively, while the reciprocal space planes $hhl$ are obtained by reconstructing the diffraction data in a single layer defined by the L1 = (1 1 0) and L2 = (0 0 1) vectors with the origin in (2.18 -2.18 0.5), resulting in a plane parallel to the $l$ direction oriented along the $k = h - 4.36$ direction in the $hk$ plane, and with the origin in (0.7 -3.7 0.25), resulting in a plane parallel to the $l$ direction oriented along the $k = h - 3$ direction in the $hk$-plane. The latter choice for the reciprocal lattice cut allows clear simultaneous observation of satellites belonging to all the 3 sets of propagation vectors.
All the satellites show sharp line profiles along the 3 directions $h$, $k$ and $l$, which indicates that such lattice modulations have a long 3-dimensional phase coherence length for all the propagation vectors, which would imply strong coupling of the Pt layers. It should be noted that, while the appearance of the $q1$ and $q2$ satellites is preceded by diffuse scattering, no detectable diffuse scattering is observed in correspondence of the $q3$ satellites sites prior to their formation. By following the temperature evolution of the intensities and full width at half maximum of the satellites [$q’1$ + (0 -4 0)], [$q’2$ + (2 -2 0)] and [$q’3$ + (-2 3 0)] (Fig. \ref{temp_satellites}), four main temperature regimes have been identified: between 300 K and T1 = 230 K a well structured diffuse scattering corresponding to the $q1$ and $q2$ satellites positions is observed, between T1 = 230 K and T2 = 110 K the $q1$ satellites are formed and their intensity increases as the temperature decreases until 110 K, where an abrupt drop in the satellites intensity is observed. Between T2 = 110 K and T3 = 60 K the $q2$ satellites are formed, their intensity grows as the temperature decreases and they coexist with the $q1$ satellites until 60 K, where an abrupt drop in both the $q1$ and $q2$ satellites intensity is observed. Below T3 = 60 K the $q3$ satellites are formed, their intensity grows as the temperature decreases and they coexist with the $q2$ satellites, whose intensity shows a similar trend as the $q3$ satellites. The values for T1 and T2 are chosen as the onset of the falling edge of the FWHM temperature dependence for the $q1$ and $q2$ satellites, i.e. when the peaks become sharp. All these periodic distortions of the lattice seem to have a different nature, which suggests that the origins of their formation are different. The $q1$ satellites are characterized by incommensurate propagation vectors with zero component along the $c$-axis, they are connected to the corresponding nesting wave-vectors $Q$ = (1/3 0 0) which, according to previously reported theoretical calculations \cite{kim2015mechanism}, modulate the Fermi surface in LaPt$_2$Si$_2$ by creating gaps at these positions resulting in the CDW state. The $q2$ satellites are characterized by propagation vectors with incommensurate components in the $ab$-plane and a commensurate component along the $c$-axis. The fact that the $q2$ propagation vectors can be expressed as linear combinations of the $q1$ vectors as:

\begin{eqnarray}
 q\prime2 & = \frac{q\prime1 + q\prime \prime1 + (001)}{2},\\ 
 q\prime \prime2 & = \frac{q\prime1 - q\prime \prime1 + (001)}{2}
\label{eq2}
\end{eqnarray}

seems to indicates that the $q2$ satellites are not due to an independent CDW $Q$ vector, but are rather related to the $q1$ satellites. The appearance of the $q2$ satellites is also accompanied by the formation of an out-of-plane commensurate modulation of the cell, probably needed to compensate the structural instability induced by the $q1$ modulation. The $q3$ satellites are characterized by propagation vectors which are linearly independent from $q1$ and $q2$, but their coordinates in the reciprocal space can be expressed in terms of rational fractions of the reciprocal lattice vectors as:

\begin{eqnarray}
 q\prime3 & = (\frac{10}{3}(110) + \frac{1}{4}(001)),\\
 q\prime \prime3 & = (-\frac{10}{3}(110) + \frac{1}{4}(001))
\label{eq3}
\end{eqnarray}

resulting in a superstructural cell which is quadrupled along the $c$-axis. The $q3$ super-cell seems to be just a consequence of the structural instabilities introduced by the incommensurate modulations of the unit cell, and it does not seem to be related to any additional CDW nesting vector. However, the very weak intensity of the $q3$ satellites, makes the refinement of the $q3$ propagation vectors very difficult. Diffraction measurements at temperatures lower than 24 K, where the intensity of the $q3$ satellites is expected to be enhanced, would be needed for precise determination of their position. 

The CDW in this material was previously reported to show indications of a 2 dimensional nature and proposed to develop in the P1 layer only. However, our observation of 3-dimensional super-lattice distortions and the long range correlation between the Pt layers seems to point towards a more uniform propagation of the CDW in LaPt$_2$Si$_2$ which probably develops also in the P2 layer.

\begin{figure}[ht]
  \begin{center}
    \includegraphics[scale=0.5]{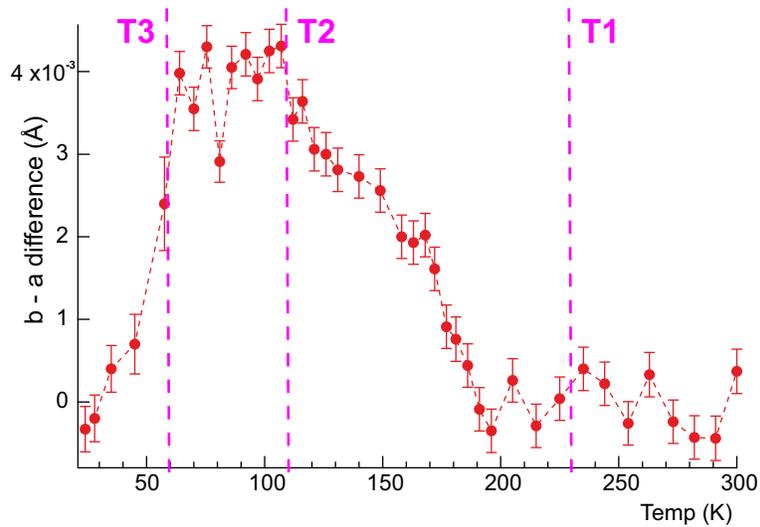}
  \end{center}
  \caption{Temperature dependence of the difference between the $b$-axis and $a$-axis in the full temperature range. The evolution of the lattice distortions is clearly visible and the transition temperatures T1, T2 and T3 are explicitly marked.}
  \label{b-a}
\end{figure}

Regarding the evolution of the crystal structure, a change in the lattice parameters occurs as a function of temperature so that the $b$-axis becomes gradually longer than the $a$-axis. Figure \ref{b-a} displays the temperature evolution of the difference $b - a$. Just like for the satellites intensities, also in this case four temperature regimes can be identified: between 300 K and T1 = 230 K $b - a$ is close to zero within the error bars and the crystal structure of LaPt$_2$Si$_2$ can be reliably refined with the tetragonal space group $P4/nmm$ (no. 129), between T1 = 230 K and T2 = 110 K the unit cell is gradually distorted on cooling as the difference $b - a$ grows. In this temperature range the $q1$ lattice modulation is incompatible with the tetragonal symmetry and induces a strain resulting in the $a$ and $b$ axes being inequivalent. Refinement attempts with space group $P4/nmm$ could not be stabilized. Therefore, from a group subgroup relationship argument, the centrosymmetric orthorhombic space group $Pmmn$ (no. 59) was identified as the most probable for a structural transition from $P4/nmm$, however, refinement attempts with this and other orthorhombic space groups (namely the non centrosymmetric $Pmn2$ and $Cmme$) could not provide a stable result either. Indeed, calculations for crystal system recognition including the main unit cell and the $q1$ modulation vectors for the T = 140 K data, performed with the software Jana, provided the triclinic and monoclinic crystal systems as the only possible symmetry choices which would make the main unit cell consintent with the $q1$ modulation. However, even though in principle the proposed triclinic and monoclinic ($P2_1/m$, $P2/m$, $P2_1$, $P2$) space groups are maximal subgroups of the tetragonal $P4/nmm$, in practice the path of symmetry reduction from the parent structure to the lower symmetry space groups requires several intermediate groups. This makes the physical realization of the structural transition highly improbable, according to the Symmetry Principle. The most reasonable among the monoclinic subgroups is $P2_1/m$, which can be obtained from the aristotype through the path  $P4/nmm \rightarrow Pmmn \rightarrow P2_1/m$. However, attempts of refining the T = 140 K data with the latter monoclinic space group did not provide any stable result either. Therefore, the structural evolution in the temperature range between T1 = 230 K and  T2 = 110 K will be described as a second order transition from a commensurate tetragonal phase to an incommensurate distorted phase in which the $b$-axis is elongated with respect to the parent tetragonal structure. The expression "distorted tetragonal" will be used from now on to identify such a phase. Around a temperature T2 = 110 K a first order transition occurs between the incommensurate distorted tetragonal phase and the centrosymmetric orthorhombic $Pmmn$ space group, along with a slight increase in the unit cell volume. The refinement of the crystal structure in this phase could be reliably stabilized with the best R factor achieved for the T = 102 K data set (see table \ref{table1}). The R factor becomes progressively worse on cooling from T = 102 K to T = 64 K.
Such a structural transition is not of displacive nature, but more of a order-disorder type. Indeed, the small value of the total amplitude of the displacive distortion ($A$ = 0.0072 Å) leaves the atomic positions of the parent structure almost unchanged in the orthorhombic cell, while the occupation probabilities of the atomic sites are different (see table \ref{table1}). It should be noted that for the refinement of the commensurate main phase the satellite Bragg reflections are not taken into account, therefore the low temperature crystal structures reported here are average crystal structures. Between T2 = 110 K and T3 = 60 K the orthorhombic symmetry is maintained, in correspondence to the coexistence of $q1$ and $q2$ satellites. Below T3 = 60 K a first order transition occurs and the tetragonal symmetry $P4/nmm$ is restored. Therefore, the structural evolution of LaPt$_2$Si$_2$ can be summarized as follows on cooling: tetragonal $P4/nmm$ $\Rightarrow$ distorted tetragonal (2nd order) $\Rightarrow$ orthorhombic $Pmmn$ (1st order) $\Rightarrow$ tetragonal $P4/nmm$ (1st order).

\begin{figure}[ht]
  \begin{center}
    \includegraphics[scale=0.6]{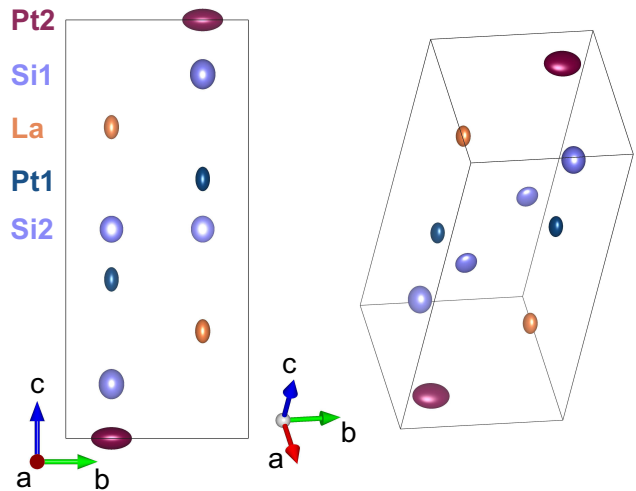}
  \end{center}
  \caption{Unit cell of LaPt$_2$Si$_2$ with the displacement ellipsoids of the respective atoms. The probability for each atom to be included in the ellipsoids was set to 100 $\%$}
  \label{biso}
\end{figure}

Table \ref{table1} reports the results of the refinement within the 3 temperature regimes [T $<$ T3], [T3 $<$ T $<$ T2] and [T $>$ T1], along with their respective reliability factors R1. The values of the R-factor for the 3 refinements, below 5$\%$, indicate a good agreement between the calculated and observed models. The refinement showed significant improvement with the adoption of an anisotropic displacement parameter $\textbf{U}$. This implies a propensity for the atoms in LaPt$_2$Si$_2$ to move away from their reference lattice positions along certain privileged directions. In particular, from the values of the diagonal elements of the tensor $\textbf{U}$ reported in table \ref{table1} (the refinement provided zero value for the off diagonal terms), the atoms Pt1 dislocate preferentially along the $c$-axis while the Pt2 within the $ab$-plane. A graphic representation of the anisotropic displacement ellipsoids extracted from the 102 K refinement results is displayed if figure \ref{biso}. This behavior seems to indicate that the Pt2 atoms are responsible for the in-plane coupling of the CDW propagation while the Pt1 atoms are responsible for the out of plane coupling along the $c$-axis. 

This fact is in contrast with the theoretical calculations and experimental evidences according to which the Pt1 layer is the only responsible for the occurrence of the CDW state in LaPt$_2$Si$_2$ \cite{kim2015mechanism, aoyama2018195pt}. Indeed our results seem to indicate that 2 CDW transitions occur, one at T = T1 = 230 K with nesting vectors $q1$, propagating in the Pt2 layer (since $q1$ has $l$ component equal to zero), and a second one at T = T2 = 110 K with nesting vectors $q2$, propagating in the Pt1 layer (since $q1$ has non-zero $l$ component). The reason for this discrepancy might be that the NMR data, which provide proof of the fact that the Pt1-5d bands are the only responsible for the CDW transition in LaPt$_2$Si$_2$ \cite{aoyama2018195pt}, were only acquired in a temperature range from 5 K to 200 K, therefore they observed the T2 transition, but not the T1. An NMR experiment in a wider temperature range might confirm or disprove this conjecture. It should be noted that a system analogous to LaPt$_2$Si$_2$ with a CaBe$_2$Ge$_2$-type structure, SrPt$_2$As$_2$, shows a double CDW transition occurring in two separate layers \cite{kawasaki2015coexistence}.

\begin{table*}[ht]
\caption{\label{table1} \textbf{Structural parameters in LaPt$_2$Si$_2$} Summary of the space groups, unit cell parameteres, atomic coordinates and sites, anisotropic Debye waller factors and reliability $R$-factors of the refinement for LaPt$_2$Si$_2$ in the three temperature regimes T $>$ T1, T3 $<$ T $<$ T2 and T $<$ T3.}
\resizebox{17.5cm}{!}{
\begin{tabular}{|c|ccccccccc|}
& & & & & & & & &\\
 T < T3 & T = 24 K & SG: $P4/nmm$ & & a = b = 4.2727 (Å), & c = 9.7987 (Å), & $\alpha = \beta = \gamma$ = 90$^{\circ}$ & & & \\
& &  & & & & & & &\\
\hline
& x & y & z   &   Occ & site, symm & U$_{11}$ & U$_{22}$ & U$_{33}$ & \\
\hline
Pt1 & 0.75000 & 0.75000  & 0.62009(4) &  1 & 2c, 4mm  & 0.00150(8) & 0.00150(8) & 0.00815(8) &  \\
Pt2 & 0.75000 & 0.25000  & 1.00000    &  1 & 2a, -4m2 & 0.0189(2)  & 0.0189(2)  & 0.0045(2)  &  \\
La  & -0.25000 & 0.75000 & 0.25613(6) &  1 & 2c, 4mm  & 0.0022(1)  & 0.0022(1)  & 0.0069(2)  &  \\
Si1 & 0.75000 & 0.75000  & 0.8707(5)  &  1 & 2c, 4mm  & 0.0113(11) & 0.0113(11) & 0.010(2)   &  \\
Si2 & 0.25000 & 0.75000  & 0.50000    &  1 & 2b, -4m2 & 0.0027(7)  & 0.0027(7)  & 0.0092(14) &  \\
\hline
\textit{R1} ($\%$) &  &  &  & 4.9 & & & & & \\
\hline
& & & & & & & & &\\
 T3 < T < T2 & T = 102 K & SG: $Pmmn$ & & a = 4.2981 (Å),  & b = 4.3021 (Å), & c = 9.876 (Å), & $\alpha = \beta = \gamma$ = 90$^{\circ}$ & & \\
& &  &  &  & &  & & & \\
\hline
& x & y & z   &   Occ & site, symm & U$_{11}$ & U$_{22}$ & U$_{33}$ & \\
\hline
Pt1 & 0.75000 & 0.75000 & 0.61995(4)   & 0.5 & 2a, mm2 & 0.0052(2) & 0.0026(3) & 0.0073(3) &  \\
Pt2 & 0.75000 & 0.25000 & 1.00011(3)   & 0.5 & 2b, mm2 & 0.01195(20) & 0.0204(3) & 0.0062(3) &  \\
La  & -0.25000 & 0.75000 & 0.25612(7)  & 0.5 & 2a, mm2 & 0.0057(2) & 0.0028(3) & 0.0071(3) &  \\
Si1 & 0.75000 & 0.75000 & 0.8703(5)    & 0.5 & 2a, mm2 & 0.010(2)  & 0.0084(22)& 0.011(2)  &  \\
Si2 & 0.25000 &  0.75000 & 0.5001(2)   & 0.5 & 2b, mm2 & 0.0027(9) & 0.0068(15)& 0.0088(15)&  \\
\hline
\textit{R1} ($\%$) &  &  &  & 4.19 & & & &  & \\
\hline
& & & & & & & & \\
T > T1 &  T = 300 K & SG: $P4/nmm$ & & a = b = 4.2793 (Å),  & c = 9.8119 (Å), & $\alpha = \beta = \gamma$ = 90$^{\circ}$ & & & \\
& &  & &  & & & & &\\
\hline
& x & y & z   &   Occ & site, symm & U$_{11}$ & U$_{22}$ & U$_{33}$ & \\
\hline
Pt1 & 0.75000 & 0.75000  & 0.61990(3)   & 1 & 2c, 4mm  & 0.0044(1) & 0.0044(1) & 0.0088(2) &  \\
Pt2 & 0.75000 & 0.25000  & 1.00000      & 1 & 2a, -4m2 & 0.0144(1) & 0.0144(1) & 0.0086(2) &  \\
La  & -0.25000 & 0.75000 & 0.25568(5)   & 1 & 2c, 4mm  & 0.0049(1) & 0.0049(1) & 0.0085(2) &  \\
Si1 & 0.75000 & 0.75000  & 0.8707(3)    & 1 & 2c, 4mm  & 0.0090(8) & 0.0090(8) & 0.011(1)  &  \\
Si2 & 0.25000 & 0.75000  & 0.50000      & 1 & 2b, -4m2 & 0.0067(5) & 0.0067(5) & 0.008(1)  &  \\
\hline
\textit{R1} ($\%$) &  &  &  & 4.3 & & & & & \\

\end{tabular}
}
\end{table*}

\begin{figure}[ht]
  \begin{center}
    \includegraphics[scale=0.7]{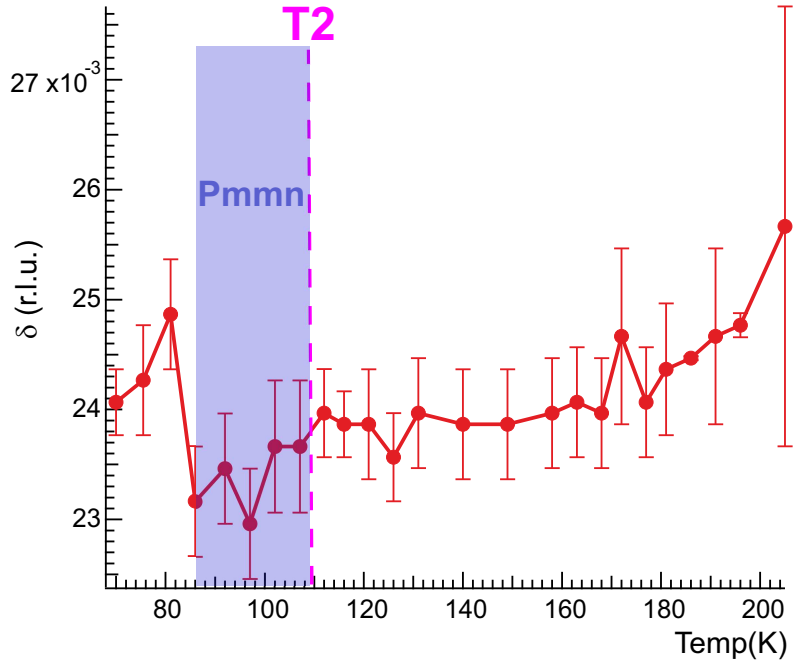}
  \end{center}
  \caption{Incommensurability of the $q\prime1$ satellites with respect to the main lattice expressed in reciprocal lattice units. The solid line is a guide to the eye.}
  \label{delta}
\end{figure}

The structural instability associated to the formation of CDW eventually results in a structural transition from distorted tetragonal to orthorhombic symmetry in LaPt$_2$Si$_2$, where ion displacements are needed to stabilize the charge perturbation. Ionic relocation requires strong electron-phonon coupling, which implies that the Fermi surface nesting/gapping cannot be the only driving mechanism involved in this transition, denoting the fact that the CDW established in LaPt$_2$Si$_2$ cannot be described within the simple Peierls picture. A normal-to-incommensurate structural transition is considered to be strong evidence of CDW formation and it was indeed observed in the prototypical 2 dimensional layered CDW compound 2H-TaSe$_2$ \cite{moncton1975study} as well as in other systems that exhibit 2 dimensional and 3 dimensional CDW \cite{wilson1975charge} \cite{hwang1983transformation}. The phenomenology in these cases foresees an alteration of the normal crystalline periodicity of the material caused by the CDW wave vector which, being determined by the Fermi surface nesting, is not necessarily an integral fraction of the reciprocal lattice vector. In this way an incommensurate structure is established in the crystal, as a result of the distortion of the parent commensurate structure, until the lattice undergoes a second distortion towards a "locked-in" commensurate structure, whose reciprocal lattice vector is an integral multiple of the CDW wave vector. Such structural changes are expected to be second order (normal-to-incommensurate) and first order (incommensurate-to-commensurate) respectively, according to the Landau theory of phase transitions, as demonstrated by McMillan in 1975 \cite{mcmillan1975landau}.
In LaPt$_2$Si$_2$ however, although we observe the occurrence of the second order normal-to-incommensurate and first order incommensurate-to-commensurate structural transitions (tetragonal $P4/nmm$ - distorted tetragonal - orthorhombic $Pmmn$), the lowest energy state with a stable structure does not seem to be achieved in this system within the investigated temperature range. Indeed, an additional first order structural transition occurs below T = 60 K. By refining the unit cell parameters together with the incommensurate $q1$ vectors across the temperature range where the $q1$ satellites are clearly visible (70 K $<$ T $<$ 205 K), it is possible to plot the degree of incommensurability $\delta$, defined as $q\prime1$ = ($\frac{1}{3} + \delta$ 0 0), for the position of the $q\prime1$ satellites relative to the main lattice as a function of temperature (Fig. \ref{delta}).

The value of $\delta(T)$ reaches a minimum between 110 K and 85 K, which corresponds to the temperature range in which the structural refinement with the orthorhombic unit cell for the main phase shows the best reliability factors (highlighted by the shaded region in figure \ref{delta}), but it is never equal to zero. Indeed the $\delta(T)$ value below 85 K increases again, in correspondence to the increase in intensity of the higher order satellites $q2$ (see figure \ref{temp_satellites}). Although a structural transition to a crystalline commensurate phase can be stabilized in the temperature range 85 K $<$ T $<$ 110 K, it cannot be labelled as a "locked-in" transition because the CDW wave vectors are never commensurate to the main phase, and indeed an additional structural transition occurs at lower temperature.
To interpret such a behavior the concept of discommensuration (DC) might be relevant in this case. DC was initially introduced theoretically by McMillan to explain the properties of 2H-TaSe$_2$ \cite{mcmillan1976theory}; here the incommensurate phase is regarded as a defect melting transition in which narrow domain walls separate large commensurate domains and, within such narrow domains, the superlattice phase fluctuates rapidly. Later on this conjecture was confirmed with a dark-field electron microscopy experiments \cite{chen1981direct}, which provided direct observation of the commensurate and incommensurate domains and demonstrated that the CDW dislocation is the main responsible of the normal-to-incommensurate transition in 2H-TaSe$_2$. In the case under investigation we have no direct evidence of DC. However, the metastable subtle structural changes in the main crystalline phase and the appearance of the higher order satellites $q2$ related to the $q1$ are experimental indications in support to the thesis that the CDW in LaPt$_2$Si$_2$ undergoes discommensuration, in qualitative agreement with the 2H-TaSe$_2$ case. Moreover, there are clear satellite intensity fluctuations in the full investigated temperature range (see figure \ref{temp_satellites}), which can be interpreted as due to coherent interference from the ordered commensurate domains in the superlattice phase, in a similar way as it was done for monolayer Kr/graphite thin film systems \cite{stephens1979x}.

\subsection{\label{sec:bulk} Bulk measurements}
\begin{figure}[h!]
  \begin{center}
    \includegraphics[scale=0.5]{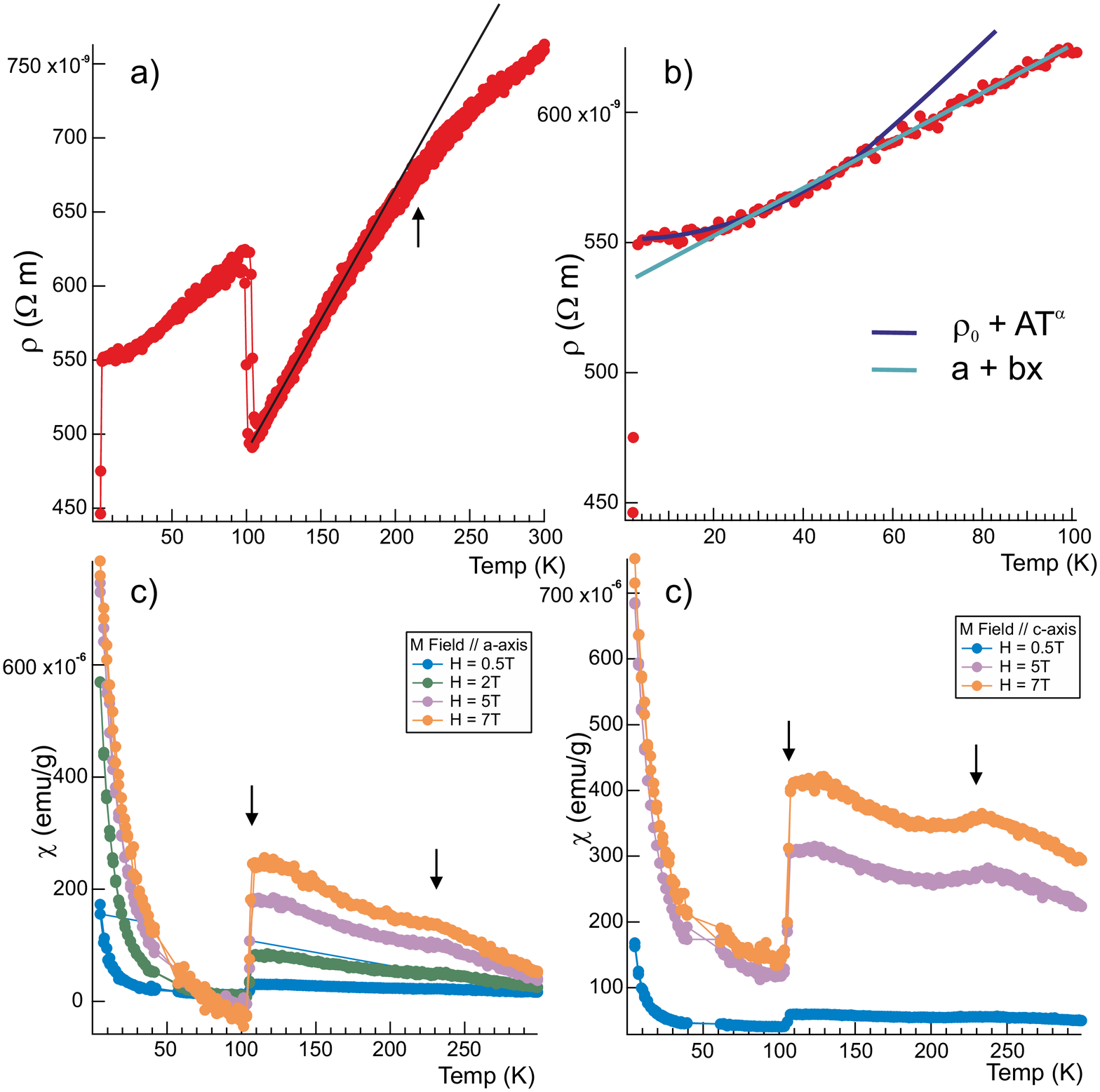}
  \end{center}
  \caption{a) Resistivity curve as a function of temperature for LaPt$_2$Si$_2$ single crystal with the electric current applied along the  $c$-axis, the solid line is a guide to the eye to highlight the change of slope in the resistivity curve b) Detail of the resistivity curve in the temperature range from 2 K to 100 K. The solid lines are fits to the power law and linear functions respectively. c) Susceptibility curve for LaPt$_2$Si$_2$ with the magnetic field flux lines oriented along the $a$-axis and d) $c$-axis with different values of the field magnitude.}
  \label{bulk}
\end{figure}

Magnetic susceptibility with magnetic field, equal to 0.5 T, 2 T, 5 T, and 7 T, applied along $a$-axis and $c$-axis and resistivity data with the current running along $c$-axis on heating and cooling are displayed in figure \ref{bulk}.
The sharp drop in the resistivity curve below 2 K is due to the onset of the superconducting transition, reportedly occurring at T$_c$ = 1.22 K. The curve between 2 K and 100 K is well fitted to a power law below 50 K and to a line above, as usually occurs in normal metals. The value of the exponent as a result of the power law fitting is $\alpha$ = 2.1 $\pm$ 0.1, which is signature of normal Fermi liquid behavior and no strong electron-electron interactions are in place. This could be indication of the fact that the superconducting state established in LaPt$_2$Si$_2$ is of conventional nature, however deeper analysis in needed to unambiguously support such a statement.
A sharp upturn in resistivity is observed in proximity of T2, across the first order incommensurate to commensurate phase transition, and indicates a gap opening at the Fermi surface. Usually in simple Peierls systems this transition occurs between a metallic and an insulating phase. However, metallic behavior below the transition suggests that the gap opening is partial and does not take place over the entire Fermi surface. Interestingly, the resistivity above T2 deviates from linearity bending downwards, with a phenomenology similar to NbSe$_2$ \cite{naito1982electrical}, in correspondence to the appearance of the $q1$ satellites across the second order commensurate to incommensurate phase transition. A possible interpretation for the negative curvature of the resistivity is that it is probably caused by local fluctuations of the CDW causing the opening of gaps on the Fermi surface which are too small to macroscopically affect the transport properties in LaPt$_2$Si$_2$.

A first order drop in the value of magnetic susceptibility is observed in proximity of T2, indicating reduction in the density of states due to partial opening of a gap at Fermi surface. Differences between data with magnetic field applied along $a$ and $c$ axes imply anisotropy of the gap. A small anomaly in the susceptibility is observed in proximity of T1, probably indicating the aforementioned small gap opening in correspondence to the appearance of the $q1$ satellites.
The published literature has not discussed this second transition in the susceptibility data as well as the negative curvature of the resistivity above T2, and attributes the first order transition at T2 to the CDW transition. With the results reported in this paper it is now possible to re-interpret these bulk measurements by placing the CDW induced gap opening in correspondence to the small anomaly in the susceptibility at T1 and by associating the sharp transition at T2 to a change in the electronic states of LaPt$_2$Si$_2$ induced by the first order structural transition from distorted tetragonal to orthorhombic crystal symmetry associated with the second CDW transition $q2$ at T2.

\section*{Discussion}

The structural evolution and the temperature dependence of the development of the density wave state in the CDW superconductor LaPt$_2$Si$_2$ was clarified with synchrotron XRD and bulk characterization measurements. From our investigation we concluded that the onset of the charge ordering is to be placed well above room temperature. Phonon dispersion measurements would be relevant to clarify this temperature dependence. On cooling, lattice distortions with multiple propagation vectors occur and their temperature evolution can be followed through the intensities of the corresponding satellites. In particular, four temperature regimes can be identified:

\begin{itemize}
  \item T $>$ T1 = 230 K: diffuse scattering is present in the $ab$-plane while broad satellites are already visible along the $c$-axis. The crystal structure of LaPt$_2$Si$_2$ can be reliably refined with the tetragonal space group $P4/nmm$.
  \item T2 $<$ T $<$ T1: satellites with propagation vector $q\prime1$ = [0.36 0 0] and $q\prime \prime1$ = [0 0.36 0] become sharp and their intensity increases on cooling. The crystal structure in LaPt$_2$Si$_2$ undergoes a second order structural transition from the room temperature commensurate tetragonal phase to a incommensurate distorted tetragonal phase.
  \item T3 $<$ T $<$ T2: satellites with propagation vector $q\prime2$ = [0.18 0.18 0.5] and $q\prime \prime2$ = [0.18 -0.18 0.5] become sharp and their intensity increases on cooling. The crystal structure in LaPt$_2$Si$_2$ undergoes a first order structural transition at T = T2 = 110 K from the incommensurate distorted tetragonal phase to a commensurate orthorhombic phase with space group $Pmmn$.
  \item T $<$ T3: satellites with propagation vector $q\prime3$ = [0.3 0.3 0.25] and $q\prime \prime3$ = [-0.3 -0.3 0.25] appear and their intensity increases on cooling from T3 = 60 K. The tetragonal structure with space group $P4/nmm$ is restored after a first order transition.
\end{itemize}

A first CDW transition should be associated to the $q1$ satellites, the $q2$ satellites, associated to a second CDW transition, are related to the $q1$ as higher order satellites and the $q3$ modulation of the lattice is established as a consequence of the structural instabilities induced by the charge modulation. First order transitions towards commensurate crystal structures for the main cell occur in correspondence to the appearance of the $q2$ and $q3$ superstructures. The long 3D phase coherence length for the $q1$ and $q2$ propagation vectors implies strong inter-planar interactions among the Pt1 and Pt2 layers. This seems to indicate that 2 distinct CDW transitions occur: one at T1 with propagation vector $q1$ in the Pt2 layer, and one at T2 with propagation vector $q2$ in the Pt1 layer. The CDW induced ion displacement indicates the presence of strong electron-phonon coupling, while the metastable structural changes, the appearance of higher order satellites and their intensity fluctuations can be indications of discommensuration of the CDW in LaPt$_2$Si$_2$. This behavior is indeed similar to the behavior of other systems that manifest CDW discommensuration and dislocation. However, no direct evidence of DC is observed in this work, therefore, it is here presented as a speculative conjecture.
The temperature dependent behavior of the crystal structure and the evolution of the CDW induced satellites clarified in this study seem to point towards an unconventional character of the LaPt$_2$Si$_2$ CDW states, with strong coupling between the Pt layers. 

\section*{Methods}

The synchrotron X-ray diffraction measurements were performed on the P21.1 beamline \cite{dippel2020examples} at the PETRA III synchrotron facility of the DESY national research center (Deutsches Elektronen-Synchrotron). The sample was mounted on a Displex cold finger cryostat, with a T-range 10 K – 320 K. The data were acquired with a PILATUS3 X CdTe 2M detector over $360^{\circ}$ omega scans in $0.1^{\circ}$ steps. Data at 41 temperatures were collected on controlled heating in different steps depending on the range [10:10:50, 55:5:120, 130:10:150, 155:5:190, 200:10:300] K. The energy of the incoming photon beam was selected to be $\approx$ 102 keV. In the low temperature region, there is a difference of about 10 K between the temperature readout in the proximity of the sample and the temperature set, therefore in the following we will always refer to the sample temperature readout. The synchrotron XRD data were collected with 3 different attenuation settings: atten0 for zero attenuation, atten2 for intermediate attenuation, atten5 for maximum attenuation. Each attenuation step is achieved through 0.1 mm Tl. Due to the strong attenuation of the atten5 setting, not much information can be extracted from this dataset, therefore it was disregarded.

Preliminary in-house XRD data were collected at the Arrhenius Laboratory in Stockholm University. The resistivity and susceptibility measurements were performed at the Physical Properties of Materials laboratory at the Paul Scherrer Institute (PSI), Switzerland. The LaPt$_2$Si$_2$ sample was prepared using arc melting of high purity La, Pt and Si at the Tata Institute of Fundamental Research (TIFR), Mumbai \cite{gupta2017superconducting}.

All images involving crystal structure were made with the VESTA software \cite{momma}, the XRD data reduction and unit cell determination were carried out with the software CrysAlis$^{\rm Pro}$ \cite{crysalispro2021rigaku}, the crystal structure refinement was performed with the software SHELX \cite{sheldrick2015crystal}, the data plots were produced with the software IgorPro \cite{igor}.

\bibliography{Refs}



\section*{Acknowledgements}

We acknowledge DESY (Hamburg, Germany), a member of the Helmholtz Association HGF, for the provision of experimental facilities. Parts of this research were carried out at PETRA III and we would like to sincerely thank O. Ivashko, M. von Zimmermann and P. Glaevecke for assistance in using P21.1. Beamtime was allocated for proposal 11013546. The bulk measurements were carried out at the Laboratory for Multiscale Materials Experiments, Paul Scherrer Institut, in Switzerland. The authors wish to thank the staff of PSI for the valuable support provided during the measurements. The initial in-house XRD characterization was carried out at the Stockholm University, Arrhenius Laboratory (Department of Materials and Environmental Chemistry). The authors wish to thank A. K. Inge for the valuable support provided during the measurements, as well as A. Geresdi and N. Trnjanin, from the Chalmers University of Technology, for the enlightening discussions around unconventional superconductors.
This research is funded by the Swedish Foundation for Strategic Research (SSF) within the Swedish national graduate school in neutron scattering (SwedNess). 

\section*{Author contributions statement}

E.N. conceived the experiments. E.N., O.I., M.M., J.L., Y.M.K. conducted the experiments. E.N, I.S. analyzed the results. The samples were synthesized by Z.H. and A.T. who also conducted the initial sample characterizations. E.N. and M.M made all the figures. E.N. created the first draft.

\section*{Data availability statement}

All the data of this work are available from the corresponding authors on request.

\textbf{Competing interests} 

The authors declare no competing interests.  

The corresponding author is responsible for submitting a \href{http://www.nature.com/srep/policies/index.html#competing}{competing interests statement} on behalf of all authors of the paper.

\end{document}